# MDAgent: A Multi-Agent Framework for End-to-End Molecular Dynamics Research

Zhenyu Ma[1#], Chunyi Yang[2#], Yuyang Song[1#], Jingyi Zhu[2#], Letian Yang[1], Limei Xu[2], Min Xiao[1], Xukai Jiang[1*]

[1]National Glycoengineering Research Center, Shandong University, Qingdao 266237, China. [2]State Key Laboratory of Microbial Technology, Shandong University, Qingdao 266237, China.

[#]*These authors contributed equally to this work*

**\*Correspondence**

Xukai Jiang, Email: xukai.jiang@sdu.edu.cn

**Abstract:**

Molecular dynamics (MD) simulation is a powerful tool for studying biomolecular structural changes, molecular recognition, transmembrane transport, and functional mechanisms. However, its practical bottleneck lies not only in software operation or parameter setup, but in translating experimental questions into executable, interpretable, and reviewable computational workflows. Here, we present MDAgent, a multi-agent system for end-to-end molecular dynamics research. The system integrates problem understanding, literature-guided strategy design, simulation execution, trajectory analysis, mechanistic interpretation, and quality supervision into a unified workflow, enabling agents not only to run simulations but also to generate research-oriented computational plans and analytical reports. We further introduce a case-based learning mechanism based on Skill and Memory, which stores reusable knowledge from prior tasks, including parameter choices, operational rules, analytical logic, and problem-solving pathways, thereby supporting cross-task transfer without retraining the underlying model. Across multiple representative molecular simulation tasks, MDAgent achieved stable end-to-end performance with improved strategic adaptability, interpretability, and generalization. In an independent complex task involving conformational transitions of TMEM16F and XKR8, the system successfully completed system design, simulation, and mechanistic analysis for large membrane proteins. These results show that combining multi-agent collaboration with case-based learning can transform MD agents from workflow automation tools into scientific question-oriented computational research systems, providing a scalable framework for AI-driven automated research.

# 1. Introduction

Molecular dynamics (MD) simulation has become an important computational tool for elucidating biomolecular structural changes, molecular recognition, transmembrane transport, and functional mechanisms[1]. However, the real bottleneck in its practical application lies not only in the complexity of software operation or the tediousness of parameter setup, but more importantly in the difficulty of effectively translating experimental research topics into executable, interpretable, and reviewable computational research workflows. To address this issue, this study proposes MDAgent, a multi-agent system for end-to-end molecular dynamics research. Organized around real scientific research workflows, the system integrates scientific problem understanding, literature retrieval and strategy design, simulation task distribution and execution, trajectory analysis and mechanistic summarization, as well as result supervision and quality scoring into a unified closed loop, thereby enabling the agents not only to "complete a simulation," but also to generate computational plans and analytical reports with genuine research value for specific biological problems. Furthermore, this study introduces a case-based learning mechanism based on Skill and Memory, which organizes parameter experience, operational rules, analytical logic, and problem-solving pathways accumulated from previous tasks into reusable knowledge units, thereby achieving cross-task capability transfer and continual evolution without retraining the underlying model. Through learning from multiple representative molecular simulation tasks, MDAgent is able to stably accomplish the full workflow from system construction to result analysis, demonstrating stronger strategy adaptability, result interpretability, and cross-case generalization capability. Furthermore, in an independent and complex research problem represented by the conformational transitions of TMEM16F and XKR8, the system successfully completed large-scale membrane protein system design, simulation execution, and mechanistic analysis, validating its potential for autonomous research on real biological questions. Overall, this study shows that the combination of multi-agent collaboration and case-based learning can elevate molecular dynamics agents from "workflow automation tools" to "scientific question-oriented computational research systems," and provides a scalable new framework for AI-driven automated scientific research[1–7].

However, in real research settings, experimental scientists often find it difficult to conduct MD studies independently, and the reason is often not simply that they do not know how to enter simulation commands or invoke software scripts. The more critical obstacle is that they usually lack the ability to translate their own experimental topics into executable simulation tasks. For example, they may not know how to select appropriate simulation targets based on the scientific question, how to construct simulation systems suited to the needs of the project, how to choose the most appropriate strategy from among many methods and parameters, how to determine analysis metrics according to specific research objectives, or how to further interpret computational results such as RMSD, RMSF, hydrogen-bond occupancy, free energy curves, or conformational changes into mechanistic conclusions with biological significance. In other words, although existing automation tools can encapsulate command-execution workflows to some extent, they often fail to address the two most critical ends of scientific research: front-end problem formulation and strategy design, and back-end result interpretation and mechanistic extraction.

The essence of this gap is that MD research has never been a simple software operation workflow, but rather a scientific decision-making process that depends on professional experience and literature knowledge. When facing a specific research problem, an experienced molecular simulation researcher usually does not directly run a fixed template. Instead, they first understand the scientific background and objectives of the problem, review the relevant literature, analyze the technical routes adopted in previous studies, determine the appropriate methodological path in light of system size, timescale, and research question, then locally adjust the simulation parameters, and ultimately relate the simulation observables to evidence from structural biology, pharmacology, or biophysics in order to form interpretable research conclusions. Therefore, what truly limits the broader application of MD in experimental research settings is not whether "simulation commands can be run automatically," but whether "the scientific decision-making process can be systematically supported."

In recent years, the development of large language models (LLMs) and agent technologies has provided new possibilities for building automated systems oriented toward scientific research tasks[8–10]. LLMs have demonstrated strong capabilities in natural language understanding, task decomposition, literature summarization, tool use, and report generation, giving them the potential to participate in complex scientific workflows[11]. However, if a single general-purpose agent is directly applied to MD tasks, clear limitations still remain. First, MD research itself is a multi-stage process involving the coupling of multiple types of knowledge, including scientific problem analysis, literature retrieval, strategy design, simulation execution, quality assessment, and biological interpretation, and the capability requirements differ across stages[12–16]. Second, the reliability of a simulation strategy cannot rely solely on one-time language generation, but must instead be grounded in domain literature, historical successful cases, and methodological experience. The formation of real scientific research capability does not depend only on the memorization of static knowledge, but more on the learning, comparison, reuse, and transfer of a large number of concrete cases. When facing a new problem, researchers often actively refer to similar previous studies: what simulation strategies were adopted, why they were designed in that way, which parameter settings were effective, which failures should be avoided, and how the final results corresponded to experimental phenomena.

In summary, what is currently lacking in the field of molecular dynamics is not merely an automation tool that can execute simulation commands or generate input files, but rather an end-to-end intelligent research framework capable of completing strategy design, task execution, result interpretation, and quality control around real biological questions. To address this issue, this study proposes MDAgent, a multi-agent system oriented toward the full workflow of molecular dynamics research. Starting from scientific questions, the system integrates literature-driven strategy formulation, simulation task distribution and execution, trajectory analysis and mechanistic summarization, as well as result supervision and scoring into a unified closed loop, and further introduces a case-based learning mechanism based on Skill and Memory, enabling the system to accumulate experience, reuse effective strategies, and improve cross-task transfer capability over continuous tasks[17]. We trained MDAgent on multiple representative molecular simulation cases, and validated its potential for autonomous research on real biological questions through an independent and complex task represented by the conformational transitions of TMEM16F and XKR8. The results show that MDAgent can not only stably complete the full process from problem understanding to research report generation, but also demonstrate good adaptability, interpretability, and scalability under complex systems and diverse scientific objectives. These findings indicate that the combination of multi-agent collaboration and case-based learning provides a feasible path toward building truly scientific question-oriented automated molecular dynamics research systems.

## 2. Results

### 2.1 Overall Architecture of MDAgent

We first constructed MDAgent, a multi-agent system for end-to-end molecular dynamics research. The system organizes a complete computational research task into a continuous scientific workflow, from scientific problem understanding, literature retrieval, and strategy formulation to simulation execution, result integration, and report quality evaluation, thereby forming a unified task closed loop (Figure 1a). This design allows the system to no longer require users to provide a complete simulation strategy in advance, but instead to autonomously develop a research plan around a specific scientific question and output interpretable research results, rather than merely raw trajectories or several generic analysis metrics.

On this basis, we further introduced a case-based learning mechanism, enabling the system's capabilities to accumulate continuously across different tasks. After completing each MD case, the system distills information such as problem type, literature retrieval path, strategy design logic, key parameter settings, analytical focus, and output quality into reusable experience, which is then aggregated into a shared experience repository for retrieval and reuse in subsequent similar tasks (Figure 1b). Therefore, the capability of MDAgent does not stem solely from one-time language generation by the underlying large model, but from its ability to gradually form transferable task experience across continuous cases. Furthermore, this case-based learning does not rely on retraining model parameters, but is achieved through the dynamic injection of external memory and skill at runtime. Specifically, the system first filters candidate knowledge according to the system type, research objective, method family, and analysis keywords of the current task, and compresses a small amount of the most relevant information into an expert packet tailored to the current task. Among these, memory mainly consists of task background, transferable design principles, failure modes, and applicability boundaries abstracted from historical cases, and its role is to help the model determine which previous problems are similar to the current one and which experiences are worth drawing upon. Skills, by contrast, are mainly composed of procedural knowledge, including method selection criteria, quality control checklists, execution gating rules, interpretation boundaries, and common error-avoidance strategies, and their role is to help the model clarify how the current task should be organized and executed, how it should be checked, and how it should be interpreted. In addition, case-based learning enhancement does not indiscriminately stack all knowledge into the model input, but instead achieves targeted transfer and constraint through the strategy of "task-based filtering, stage-based injection, and boundary-based use." At the same time, this runtime knowledge-injection design makes the capability enhancement of MDAgent relatively decoupled from the underlying execution framework: as long as a new agent platform supports similar memory/skill mounting and structured workspace mechanisms, the relevant experience units can in principle be transferred and reused, thereby improving the methodological portability and reproducibility potential of the system.

In addition to the main workflow and the case-based learning mechanism, MDAgent also adopts a clearly defined multi-agent division-of-labor structure (Figure 1c). Among these, the literature and strategy-design agent is responsible for integrating the literature around the scientific question and generating the simulation plan, the simulation agent is responsible for completing system construction, molecular dynamics simulation, and basic analysis according to the plan, the reporting agent is responsible for summarizing the results and generating a structured scientific report, and the review agent examines and scores the output quality from an independent perspective.

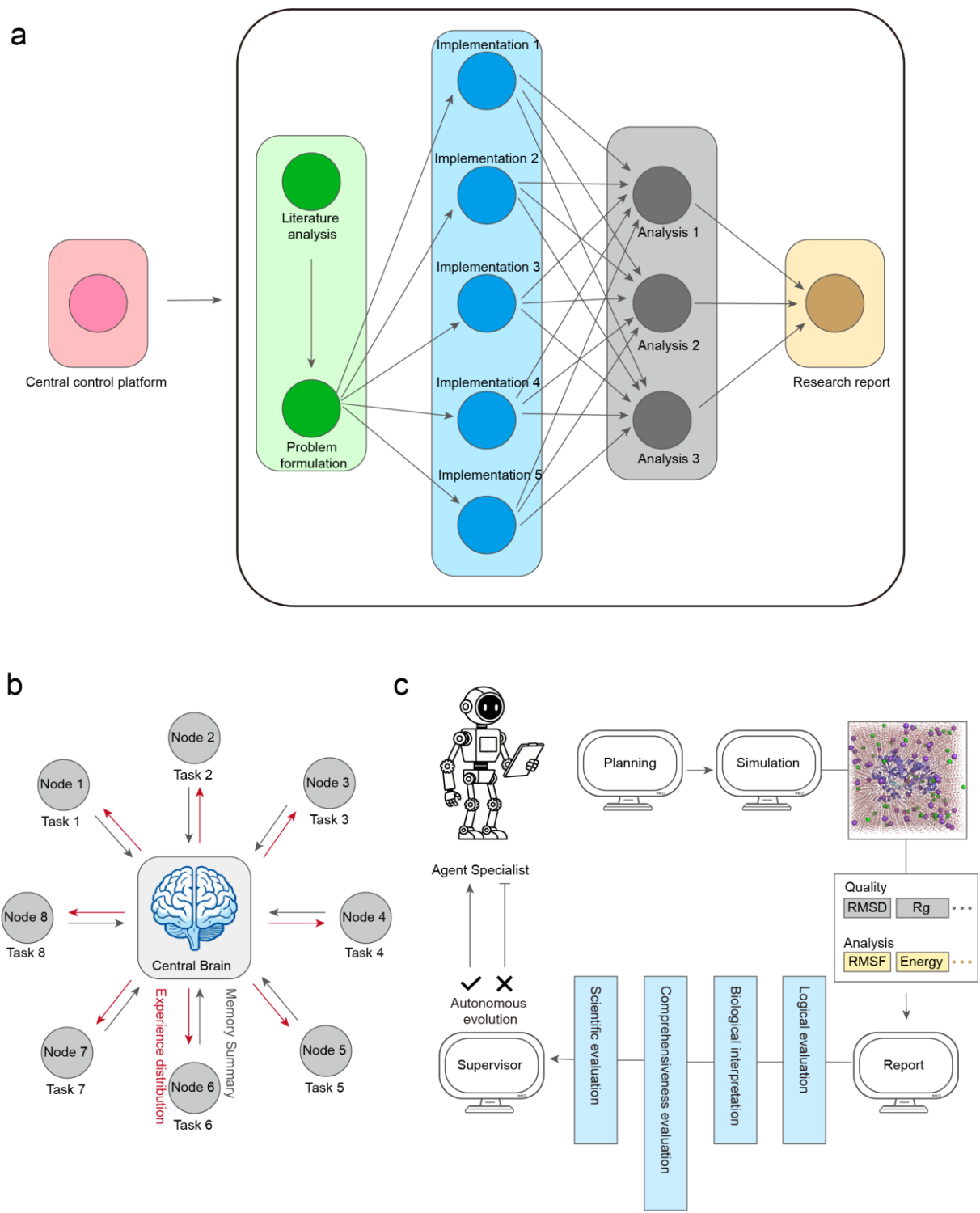

**Figure 1. Overall architecture of the MDAgent framework for end-to-end molecular dynamics research.** (a) Schematic illustration of the main workflow of MDAgent. Starting from a scientific question proposed by the user, the system sequentially performs literature

retrieval and integration, generation of simulation strategies and parameter recommendations, system construction and molecular dynamics simulation execution, basic analysis, and final scientific report generation. (b) Case-based learning mechanism of MDAgent. After completing each MD case, the system summarizes information such as problem type, literature retrieval path, strategy design logic, key parameter settings, analytical focus, and output quality into reusable experience units, which are then aggregated into a shared experience repository for retrieval and reuse in subsequent similar tasks, thereby enabling continual accumulation of task experience and cross-task transfer. (c) Multi-agent division-of-labor structure of MDAgent. Different agents are respectively responsible for literature retrieval and strategy design, simulation execution, report generation, and result scoring and review. Under a unified framework, this architecture achieves functional decoupling of strategy design, simulation execution, result interpretation, and quality control.

**2.2 Case-based learning progressively builds a fully capable and biologically grounded MD agent**

To enable the multi-agent framework to progressively acquire molecular dynamics research capabilities approaching those of domain researchers over continuous tasks, we constructed nine representative end-to-end MD cases, covering typical problems including standard soluble-protein simulation, mechanistic analysis of protein–ligand complexes, protein–DNA complex studies, all-atom comparison of GPCR active and inactive states, coarse-grained identification of GPCR lipid environments, enhanced-sampling-based free energy interpretation, evaluation of small-molecule transmembrane permeability, gating analysis of transport proteins, and multicomponent ultra-large membrane protein complexes. In these nine cases, the molecular objects, simulation systems, computational protocols, and analysis workflows were all autonomously designed and executed by the agents around specific biological questions (Figure 2). For each case, the multi-agent system completed the full process from literature understanding, system construction, parameter selection, and simulation execution to result interpretation, and extracted the key issues exposed during the process into reusable case memory and operational skills, thereby enabling the MD agent to grow from basic execution capability into a learning process that can "design systems around biological questions, judge methodological boundaries, and generate biologically grounded reports."

The starting point of case-based learning is to first enable the agent to truly perform a standard MD task correctly. The aqueous simulation of 1AKI lysozyme under the CHARMM36 force field provided the system with the most basic yet also the most critical training scenario. In this case, the agent first mastered the basic organization of a standard soluble-protein system, including force field and water model selection, energy minimization, NVT/NPT equilibration, and the sequential execution of production MD. At the same time, the system also accumulated engineering experience during real execution, including environment compatibility fixes, GPU parameter optimization, and GROMACS version switching. This case enabled the system to clearly establish the methodological rule of "PBC correction first." Uncorrected trajectories produce obvious across-box artifacts, which further distort key metrics such as RMSD; only after trjconv -pbc nojump -center processing did the Cα RMSD of lysozyme stabilize at 2.47 ± 0.66 Å, with the Rg remaining at 18.30 ± 0.28 Å, and the free energy landscape showing a single stable basin. In this way, the agent progressed from "being able to call commands" to "being able to complete a standard simulation task while ensuring the reliability of the basic results."

This foundational capability was further reinforced in the p53–DNA complex case. Focusing on the binding of the wild-type p53 DBD to the DNA response element, the system autonomously constructed a complex system containing protein, DNA, ions, and solvent environment, and completed a 200 ns simulation together with interface analysis. The results showed that the system remained in a stable bound state for 83.7% of the simulation time, with an RMSD of 3.09 ± 0.49 Å in the stable state and an average of 21.3 ± 3.0 protein–DNA hydrogen bonds. This case further trained the system in PBC correction, structural alignment, hydrogen-bond analysis, and contact analysis for complex systems, and also enabled it to recognize that "being able to simulate" means not only being able to run the system to completion, but also being able to obtain reliable trajectories and correctly carry out post-processing.

After mastering the basic execution workflow, case-based learning further enabled the agent to acquire higher-level capabilities, namely, no longer treating simulation systems and computational strategies as fixed templates, but instead proactively designing appropriate modeling schemes and sampling methods around specific biological questions. The HIV-1 protease–DMP450 case first made the system clearly recognize that system design must serve the mechanistic analysis of catalytic residues, the active pocket, and flexible functional regions, rather than remaining at the level of simple protein–ligand assembly. The NTSR1 lipid fingerprint case then extended this capability to membrane proteins and cellular membrane environments, teaching the agent that when membrane receptor–lipid interactions are involved, it must not only consider membrane composition and lipid identity themselves, but also further determine when a coarse-grained model is needed to represent membrane-environment reorganization over longer timescales and larger spatial scales. The IAA blood–brain barrier permeation case further gave the system the ability to combine methods according to the problem. For the task of transmembrane transport, the system was able to jointly design SMD and umbrella sampling as complementary tools: using SMD to generate a reasonable transmembrane pathway and identify key interfacial regions, and then using umbrella sampling to reconstruct the transmembrane free energy profile, resolve barrier locations and interfacial preferences, and interpret the transmembrane mechanism in combination with molecular physicochemical properties and membrane-environment characteristics. On this basis, the GLUT1 transporter case further showed that the system was able to transfer this method-combination capability to the study of more complex transport mechanisms, simultaneously characterizing nonequilibrium pulling responses and near-equilibrium free energy changes, no longer relying on single-path analysis, but instead being able to describe the substrate transport process more systematically from both kinetic and energetic perspectives. The Melatonin–MT1 case taught the system that when the research objective is the pathway of ligand entry into the receptor binding pocket, the residence states, and the free energy landscape, metadynamics should be preferentially adopted to track pathways and state transitions. The D2R REMD case further trained the system to proactively introduce replica exchange molecular dynamics in problems involving conformational-state comparison, so as to more appropriately sample the conformational distributions of the active and inactive states. In addition, the system also learned to make appropriate choices among different strategies, including standard all-atom MD, coarse-grained simulation, metadynamics, umbrella sampling, SMD, and REMD, according to the problem type, and further to design wild-type, control-state, and even mutant systems, so that the simulation strategy truly serves the biological question itself.

The next role of case-based learning is to gradually transform the agent from a caller of generic metrics into a mechanistic interpreter that organizes analysis around scientific questions. In the HIV-1 protease–DMP450 case, the system recognized that using only generic metrics such as RMSD, RMSF, and Rg was insufficient to answer the question of inhibitory mechanism, and therefore reorganized the analytical chain around ligand RMSD, the key Asp25–ligand distance, hydrogen-bond occupancy, and flexibility changes in the flap region, thereby elevating numerical results into the mechanistic interpretation that "the ligand locks the catalytic site and restricts flap dynamics." In the NTSR1 case, the system no longer treated RDF simply as a statistical curve, but further connected lipid binding with receptor surface electrostatics and G-protein coupling state, thereby proposing a state-dependent lipid recruitment model. Similarly, in the GLUT1 case, the analytical focus was no longer merely on comparing barrier heights, but was further organized around how the gating network controls transport: Q161–W388–H160 constitutes the key structural switch of intracellular gating, while L276 and F245 suggest potential allosteric sites. The AdeB trimer case further advanced this problem-oriented analysis to the multi-subunit level, where the system learned to organize analysis around interchain asymmetry, coordination between inner and outer gates, and the requirement for unidirectional efflux, rather than averaging the three chains. In the D2R REMD case, the system selected the TM3–TM6 distance and TM6 displacement as highly informative metrics around the classical structural hallmarks of GPCR activation, in order to distinguish the conformational distributions of the active and inactive states; in the p53–DNA case, the analytical focus was further organized into an integrated framework involving protein–DNA interfacial hydrogen bonds, transitions between stable and unstable states, and the potential effects of cancer hotspot mutations. Case-based learning also showed that the functional mechanisms of complex assemblies must be analyzed through chain-level cooperative analysis. Around the question of "whether the three subunits of the efflux pump are symmetric and which gate controls efflux," the system first learned residue numbering mapping, unit unification, and chain-specific selection in multichain systems. Thus, what the system acquired was no longer merely the ability to "run a default analysis script to completion," but the ability to "first understand the question, and then invoke the most appropriate analytical template."

In addition to system design, simulation execution, and analytical decision-making, case-based learning also continuously trained the system to generate MD result reports that conform to the reading habits of humans, especially biologists. The HIV-1 protease–DMP450 case marked the starting point of this capability. In this case, the system clearly realized that biological researchers do not simply want to see technical lists such as "what the ligand RMSD is" or "how many hydrogen bonds are present," but are more concerned with the mechanisms implied by these values. Accordingly, the system began to write MD results in the language of structural biology, such as "DMP450 locks the Asp25 catalytic site," "the flap is restricted in a closed conformation," and "the inhibitor acts by limiting functional flexibility." Subsequently, this reporting style was continuously reinforced across multiple cases: in the NTSR1 case, the system translated RDF and lipid depletion results into a "receptor state-dependent lipid recruitment model"; in the IAA BBB case, the PMF curve was further written into the physiological interpretation that "charged metabolites are more likely to require transporter assistance to cross the blood–brain barrier"; in the AdeB case, multichain gate dynamics were organized into an "asynchronous ratchet-like unidirectional efflux mechanism"; in the D2R REMD case, TM3–TM6 distributions and effect sizes were integrated into the GPCR activation narrative that "the active-state conformational landscape is more open and more heterogeneous"; and in the p53–DNA case, this capability was further extended into a molecular pathology context, enabling the system to connect the stability of the wild-type complex and the interfacial hydrogen-bond network with the potential functional consequences of hotspot mutations. By this stage, the system no longer produced technical reports, but was instead continuously learning how to rewrite trajectory data into mechanistic results that researchers in structural biology, membrane biology, transport biology, and molecular pathology could directly understand and use.

| a | b | c |
|---|---|---|
| Lysozyme | HIV System | p53-DNA |

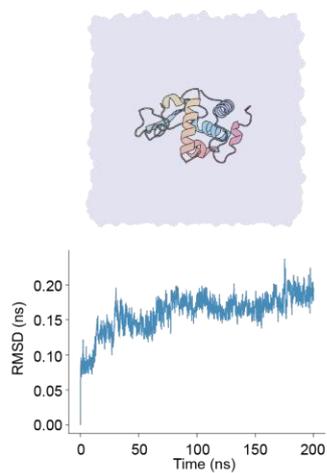 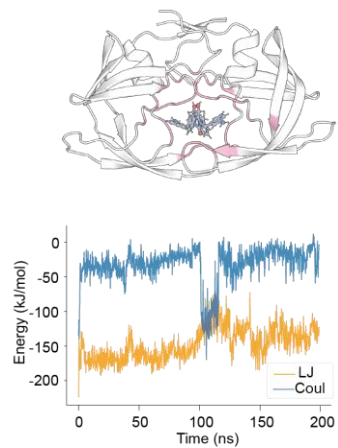 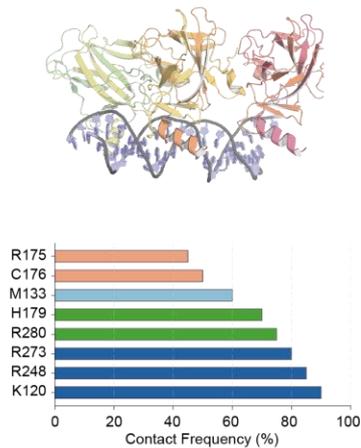

| d | e | f |
|---|---|---|
| BBB Permeation | GLUT1 Transport | CG GPCR |

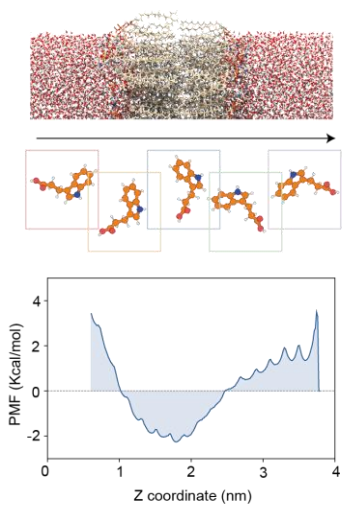 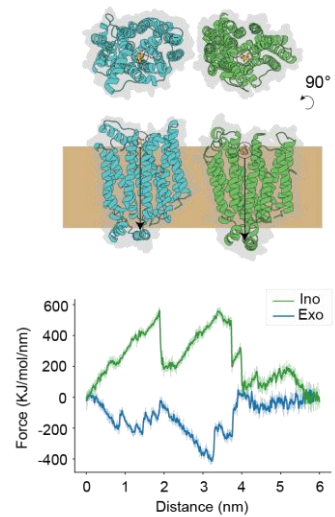 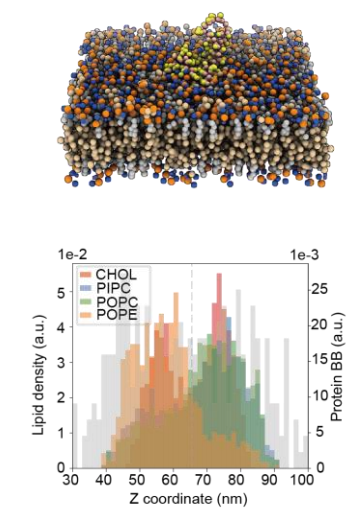

| g | h | i |
|---|---|---|
| REMD Sampling | GPCR Metadynamics | AdeB Efflux Pump |

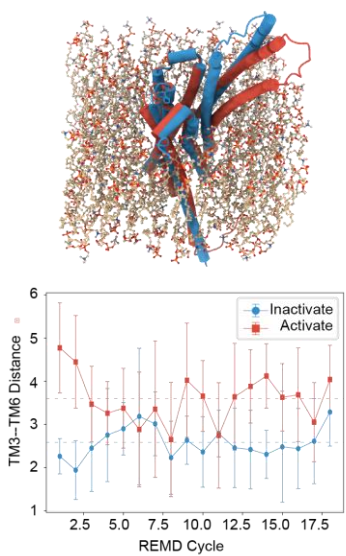 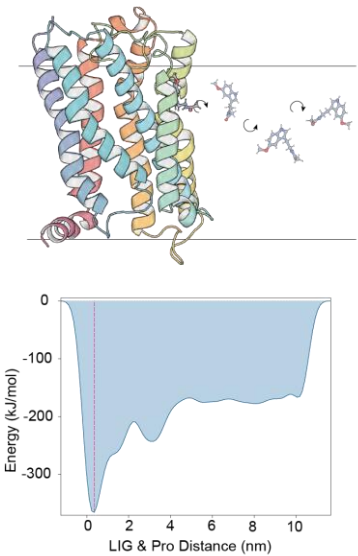 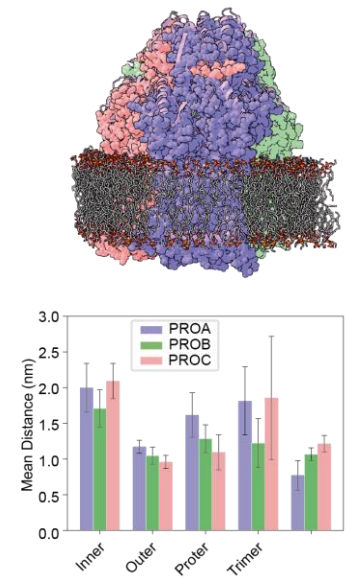

**Figure 2. Representative cases used to construct a transferable molecular simulation knowledge base, covering diverse biological systems and methodological scenarios.** (a) All-atom simulation system of lysozyme and its RMSD analysis, used to illustrate the standard MD workflow for soluble proteins and the evaluation of overall structural stability. (b) HIV-1 protease–ligand complex and interaction energy decomposition analysis, in which the Lennard-Jones (LJ) and Coulomb terms are presented separately to characterize the hydrophobic and electrostatic contributions to binding stability. (c) Structure of the p53–DNA complex and interfacial contacting amino acids, highlighting residue contacts in protein–nucleic acid recognition. (d) Schematic illustration of IAA crossing the blood–brain barrier and the corresponding PMF curve, representing the free energy analysis framework for trans-interface transport. (e) Schematic illustration of substrate pulling in two conformations of the GLUT1 transporter and the corresponding pulling-force curves. (f) Coarse-grained membrane protein system of GPCR-NTSR1 and lipid density maps, showing membrane-environment reorganization and lipid enrichment around the receptor. (g) Dopamine receptor GPCR system and conformational transition-related distance parameters, used to monitor activation-related structural changes. (h) Schematic illustration of substrate binding to the melatonin receptor and the metadynamics-derived PMF, demonstrating the capability of enhanced sampling to characterize the free energy landscape of binding/dissociation. (i) Simulation system of the multicomponent efflux pump AdeB and distance-change plots of key sites, used to characterize multidomain coupling and cooperative conformational changes.

Taken together, the nine cases show that what case-based learning confers to the system is not a set of isolated "answers," but a transferable set of molecular dynamics research capabilities that can be applied across different tasks. At the level of system design, the system is now able to proactively choose molecular objects, membrane environments, control states, enhanced-sampling pathways, and mutant schemes around biological questions, rather than relying on fixed templates. At the execution level, the system has gradually mastered a series of protocol knowledge ranging from standard soluble-protein simulation to membrane proteins, free energy calculation, multi-replica enhanced sampling, and multimeric complex analysis, including environment compatibility fixes, PBC correction, coarse-grained membrane diagnostics, RDF bulk normalization, WHAM/PMF reconstruction, metadynamics convergence inspection, SMD result interpretation, and REMD consistency handling. At the analysis level, the system has also become able to select appropriate metrics and analytical frameworks for specific scientific questions, gradually forming a problem-oriented mode of analysis spanning active-site locking, lipid fingerprints, transmembrane energy barriers and gating networks, interchain cooperativity, activation-state conformational statistics, and protein–DNA interfacial dynamics. At the level of result presentation, the system is further able to organize trajectories and numerical results into mechanistic reports that fit the contexts of structural biology, membrane biology, transport biology, and molecular pathology. More importantly, these capabilities are not simply juxtaposed, but are continuously transferred, accumulated, and superimposed across cases, allowing the system to gradually evolve from an automation tool for executing simulation workflows into an agent capable of organizing a complete MD research process around biological questions.

## 2.3 Comparative analysis of different capability combinations in new research tasks

To evaluate the performance of different system architectures in generating research blueprints for new tasks, we constructed 10 entirely new benchmark cases: KRas, CFTR, P-gp, SARS-CoV-2 Mpro, NMDA receptor, AQP4, PARP1, ANT1, TLR4–MD2, and CCR5. For each target task, all methods started from the same natural-language research description and independently generated a complete simulation plan, execution blueprint, and biological report draft. Therefore, this evaluation does not assess the model's ability to reproduce existing answers, but instead tests whether, when faced with a new research problem, it can organize a molecular dynamics study that is scientifically reasonable, executable, and equipped with a closed loop of interpretation.

We systematically compared the overall performance of five method configurations in molecular dynamics research blueprint generation, including Single-Agent LLM, Multi-Agent w/o CBL, Skill Only, Memory Only, and Full MDAgent + CBL. These five methods are different combinations of three core modules: the multi-agent collaboration module, the case memory module, and the skill module. Among these components, multi-agent collaboration enables stage-wise task decomposition, case memory supports the retrieval and transfer of prior experience, and the skill module provides procedural execution constraints and analytical guidelines (Figure 3a). The results showed that the average overall core quality of the five methods, from lowest to highest, was as follows: Single-Agent LLM (67.00), Multi-Agent w/o CBL (70.87), Skill Only (74.64), Memory Only (77.99), and Full MDAgent + CBL (87.92) (Figure 3b). These results indicate that multi-agent organization itself can bring a certain degree of improvement, and that further introducing either the case memory module or the skill module on this basis can continue to enhance the core quality of the system; only when multi-agent collaboration, case transfer, and skill constraints are all present does the system exhibit the most significant and most stable overall advantage.

The differences among the methods showed clear capability stratification (Figure 3c). Full MDAgent + CBL achieved the best average performance across all four dimensions—Plan, Execution, Report, and Adaptation & Transfer—with scores of 98.0, 81.75, 87.6, and 91.25, respectively. Among these, its lead in the Plan dimension indicates that the complete system has greater consistency in method selection, key parameter setting, and research plan organization; the improvement in the Execution dimension shows that its execution blueprints have stronger deliverability, workflow coherence, and completeness of quality control; the advantage in the Report dimension indicates that its outputs are more comprehensive in scientific correctness, depth of mechanistic interpretation, and maturity of content organization; and its marked lead in the Adaptation & Transfer dimension demonstrates that the complete system has a more stable advantage in task adaptation, cross-stage tracking, and in converting prior experience into actual incremental value for the current task. In contrast, although Multi-Agent w/o CBL already outperformed Single-Agent LLM in organization and execution workflow, its overall improvement remained limited. Although Memory Only and Skill Only each brought additional gains in experience transfer and procedural execution, respectively, these gains were still insufficient to support end-to-end system-level optimal performance. In other words, the advantage of the complete system does not arise from a local improvement in any single dimension, but from broad and coordinated enhancement across all four quality dimensions.

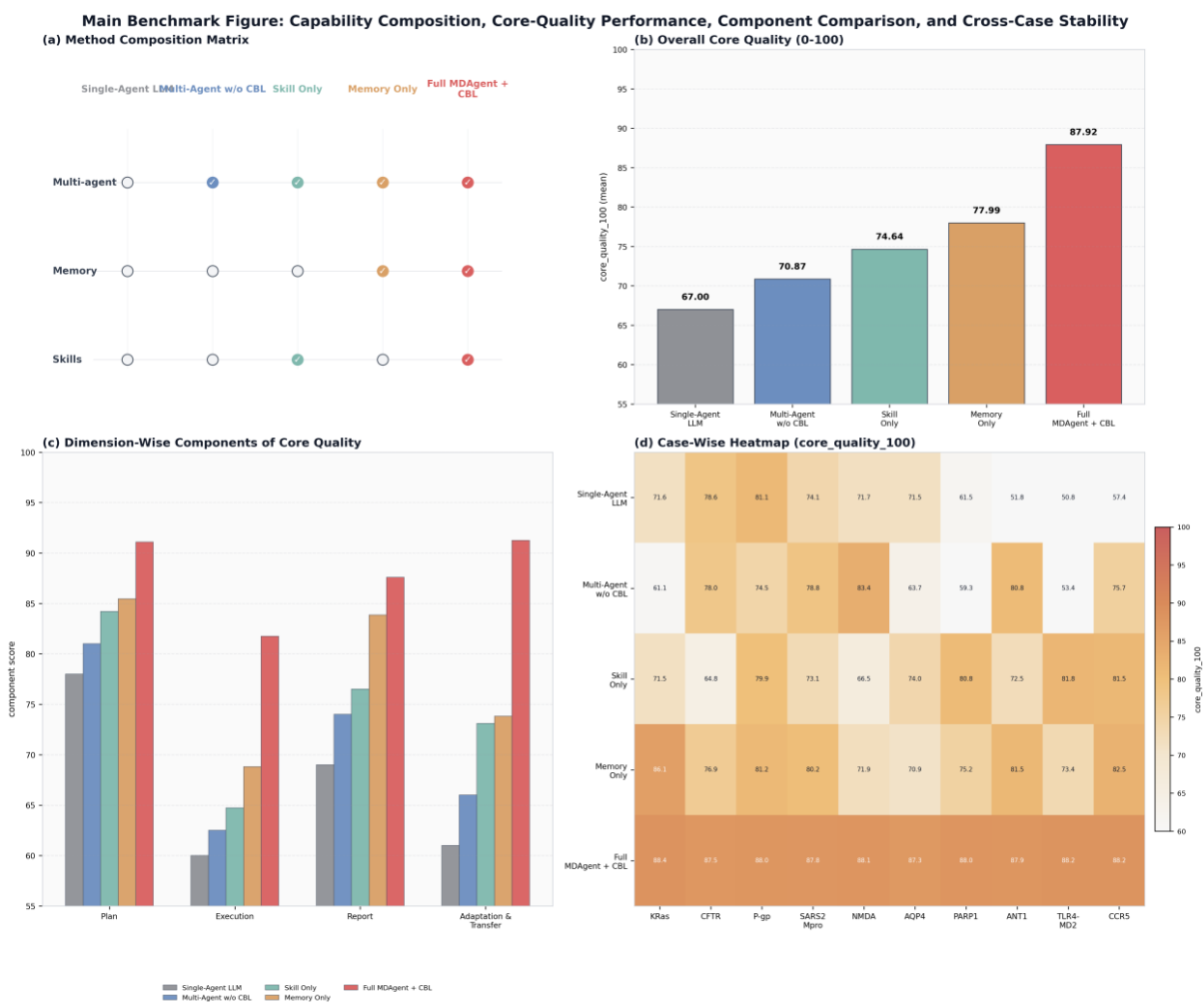

**Figure 3. Comparison of core quality among different capability combinations in the molecular dynamics research blueprint generation benchmark.** (a) Capability composition matrix of the five compared methods. From top to bottom, the rows indicate whether multi-agent collaboration, case memory, and the skill module are enabled. (b) Mean overall core quality of the five methods across 10 tasks. The average scores for each method are: Single-Agent LLM, 67.00; Multi-Agent w/o CBL, 70.87; Skill Only, 74.64; Memory Only, 77.99; and Full MDAgent + CBL, 87.92. The results show that the complete system still demonstrates the most significant overall advantage. (c) Comparison of the four primary components that constitute the total score, including Plan, Execution, Report, and Adaptation & Transfer. Full MDAgent + CBL achieved the highest mean score in all four components, reaching 91.10, 81.75, 87.60, and 91.25, respectively. At the same time, Skill Only and Memory Only also

outperformed Single-Agent LLM and Multi-Agent w/o CBL overall, indicating that both the skill module and case memory can stably improve core quality. (d) Case-wise heatmap of core_quality_100 for the five methods across the 10 benchmark tasks. This panel is used to examine whether the differences among methods remain stably present across tasks.

To further examine whether this advantage is stable across tasks rather than being driven only by a small number of high-scoring cases, we compared the case-wise performance of the five methods across the 10 benchmark tasks (Figure 3d). The results showed that Full MDAgent + CBL maintained the highest or near-highest level in most tasks, rather than exhibiting an advantage only by chance in a few individual cases. At the same time, the fluctuations of Memory Only, Skill Only, Multi-Agent w/o CBL, and Single-Agent LLM were more pronounced in several complex tasks.

Overall, this set of comparative results shows that multi-agent organization itself can improve the output quality of complex molecular dynamics research tasks, but the extent of improvement remains limited. Introducing case memory or skill knowledge alone can enhance certain capability dimensions, yet is still insufficient to support end-to-end system-level optimal performance. Only when multi-agent collaboration, case-memory transfer, and skill-based execution constraints are all present can the system achieve a truly stable comprehensive advantage in research plan completeness, deliverability, maturity of result organization, and cross-task adaptability.

**2.4 Trained MDAgent can independently complete the design and simulation of complex research tasks**

After completing system framework construction, case-based learning validation, and baseline comparison, we further applied the trained MDAgent to a complex membrane-protein research problem that was not included in the benchmark, in order to examine whether it could independently complete the full research workflow from problem understanding to mechanistic reporting in the absence of manually specified procedures. We selected two representative membrane lipid scramblases, TMEM16F and XKR8, as the research targets, and required the system to autonomously propose a research strategy, carry out simulations, generate an analytical report, and score the entire analysis process around the dynamic transition of both proteins from the closed state to the open state (Figure 4a).

After the user proposed the question of "studying the dynamic transition of TMEM16F and XKR8 from the closed state to the open state," Agent 1 (Planning & Distribution) first conducted literature retrieval and synthesis around the scramblase mechanisms of TMEM16F and XKR8. The literature analysis showed that although both classes of scramblases participate in phosphatidylserine externalization, their activation origins, structural constraints, and conformational transition pathways are not the same. TMEM16F belongs to the class of $Ca^{2+}$-activated phospholipid scramblases, and reported structures show that this protein forms a membrane-embedded dimer, in which each subunit contains a transmembrane core region associated with $Ca^{2+}$ binding and lipid transport; its functional activation is usually related to local transmembrane helix rearrangement and exposure of the lipid pathway after $Ca^{2+}$ binding. In contrast, XKR8 belongs to another class of apoptosis-related phospholipid scramblases, whose resting state is constrained by autoinhibition from the C-terminal region, and caspase-3-mediated C-terminal cleavage is considered the key triggering event that relieves inhibition, induces pathway opening, and promotes membrane lipid externalization. Based on this literature understanding, the agent further defined the research questions of the two systems in a structured manner: for TMEM16F, the core question was defined as how $Ca^{2+}$ binding drives the transition from the closed state to an open/activated conformation associated with membrane lipid scrambling; for XKR8, the core question was defined as how caspase-related C-terminal cleavage releases the inhibitory effect and induces conformational rearrangement associated with opening of the lipid transport pathway (Figure 4b). For TMEM16F, the agent selected 8B8Q as the starting simulation system. This structure is in a $Ca^{2+}$-bound closed conformation, and therefore both retains the activation-related calcium-bound state and serves as a closed-state starting point for observing subsequent conformational evolution. For XKR8, the agent proposed removing the C-terminal amino acids after D355 during modeling, in order to simulate the activated state in which the inhibitory fragment is released after caspase-3 cleavage.

After obtaining the plan, the simulation agent, Agent 2 (Perform & Experiment), translated the above research strategy into an actually executable membrane-protein simulation workflow (Figure 4c). For TMEM16F, the simulation agent used 8B8Q as the initial structure and constructed a membrane-environment system while retaining the $Ca^{2+}$-bound state. For XKR8, the simulation agent removed the C-terminal fragment after D355 during the structural modeling stage to simulate the activation-related state after caspase-3 cleavage, and then constructed the membrane-environment system. It generated the corresponding GROMACS files and ran the simulations. Following the above strategy, the agent completed system construction, membrane embedding, ion and lipid environment setup, and molecular dynamics execution, and obtained MD simulation results by running the jobs on the server.

During the result integration stage, the reporting agent, Agent 3 (Summary & Analysis), extracted the trajectories and analyzed them using GROMACS's own parameter analysis commands (Figure 4d). The agent first summarized the mechanistic differences between the two classes of scramblases based on published studies, and on this basis formulated differentiated research hypotheses. For TMEM16F, the agent proposed that if $Ca^{2+}$ binding does indeed participate in activation, then the conformational changes should first appear near the calcium-binding region and then gradually propagate through a local residue network to the pathway or terminal pocket associated with lipid passage. Therefore, the study of TMEM16F should not be limited to global RMSD analysis, but instead be structured around the mechanistic sequence of calcium-binding region rearrangement, pathway propagation, and terminal pocket response. For XKR8, the agent proposed that if C-terminal cleavage relieves the inhibitory constraint, then the cleaved state should exhibit quantifiable differences relative to the closed state in the degree of terminal pocket opening, the contact network of pathway residues, and the local interface relaxation pattern. Therefore, the focus for XKR8 should be placed on "cleavage-induced local opening."

For the TMEM16F 8B8Q system, after consulting the literature, the reporting agent identified the I515–K616 distance as the pocket marker. The results showed that this distance increased markedly, supporting a transition from a close-like state to an open-like state. For the XKR8 8XEJ system, the reporting agent defined the W45–I152 distance in chain B as the pocket marker. The results showed that this distance also increased markedly in the cleaved system, supporting a change from a close-like state to an open-like state.

In a more detailed analysis of conformational transition, the TMEM16F 8B8Q system and the XKR8 8XEJ system exhibited two clearly distinct activation propagation pathways. For 8B8Q, $Ca^{2+}$ binding first rearranged the acidic coordination network on TM6–TM8, then propagated the change to the gating layer formed by K616 on TM6 and I515 on TM4, and further induced cooperative remodeling of the pocket-wall residues Q437/I475/I520/V561/Y569 on TM3–TM5, ultimately converting the substrate transport pocket into a more open-like conformation favorable for substrate transport. In 8XEJ, after caspase-3 cleavage at the D355 site, the D355–G356 peptide bond was broken, causing the C-terminal tail segment from residues 356–382 to lose its covalent coupling with the main body, and the C-tail helix spanning residues 371–382 no longer stably constrained the cytoplasmic-side conformation. This perturbation first acted on T264/V265 at the end of TM7 and V276/A277/E278 on the TM7–TM8 linker, then propagated to L140/E141/I152 on TM4, and ultimately induced rearrangement of W45/A46/V49/L50/L53/G54 in the outer segment of TM2. The result was that W45 on TM2 moved clearly away from I152 on TM4, that is, the W45–I152 distance increased, indicating that cleavage promoted the transition of the terminal pocket from a close-like to an open-like conformation.

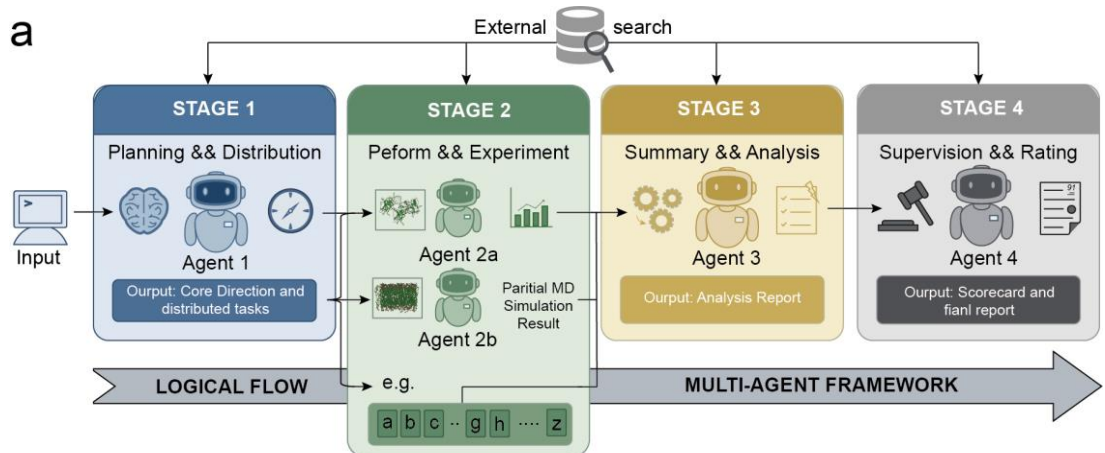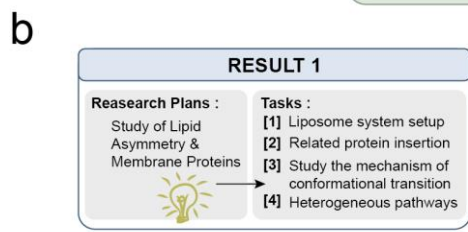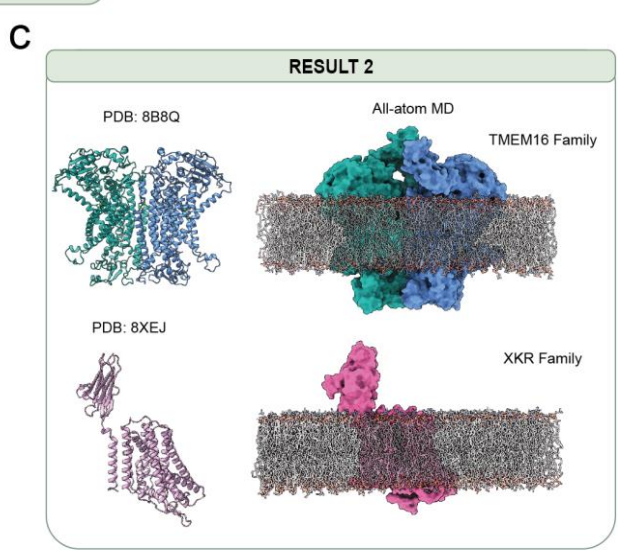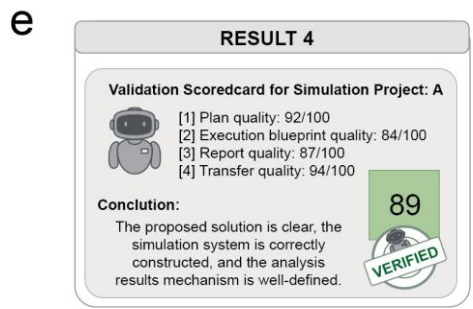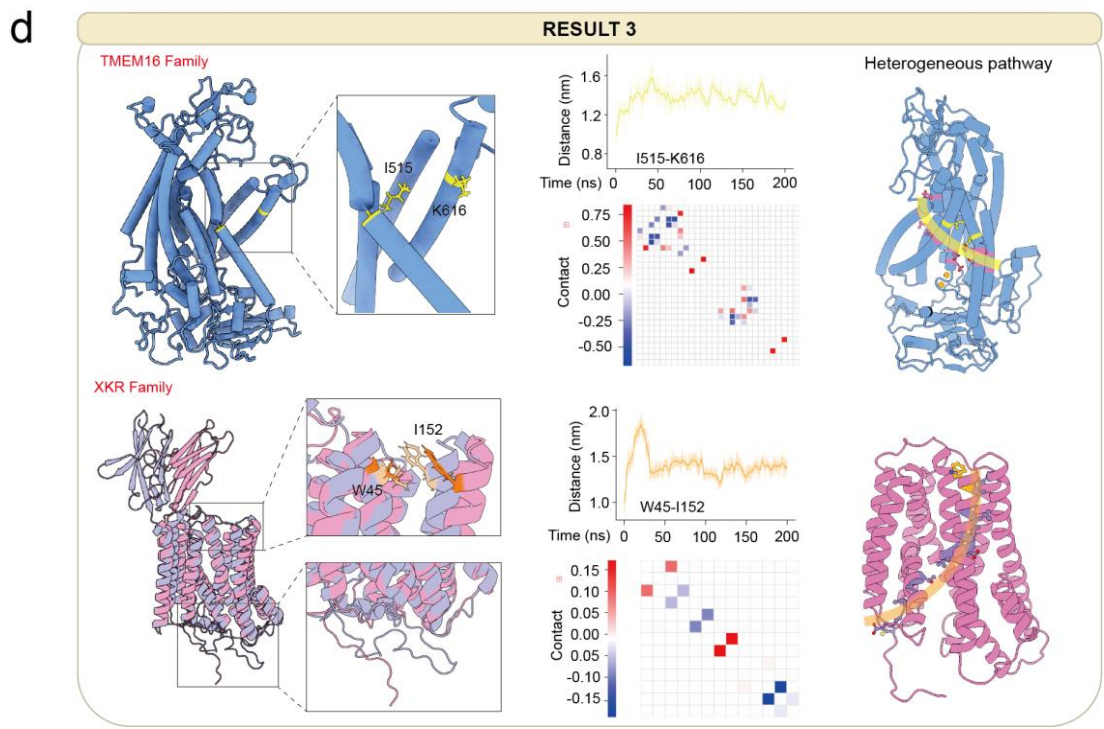

**Figure 4. Autonomous study of scramblase conformational transitions by MDAgent.** (a) Schematic illustration of the workflow of MDAgent in this independent research task. The entire workflow consists of four collaborative modules: Planning & Distribution is responsible for determining the core research direction and distributing the distributed simulation tasks; Perform & Experiment is responsible for system construction and molecular dynamics simulation execution, and outputs local simulation results; Summary & Analysis is responsible for integrating trajectory evidence and generating the analytical report; and Supervision & Rating is responsible for supervising and evaluating the entire research workflow, and outputs the scorecard and final report. (b) Output of the planning module, showing the literature-driven research strategies and task decomposition for TMEM16F and XKR8. (c) Simulation systems of TMEM16F and XKR8 constructed by the experiment module. (d) Simulation analysis results. The curves show the dynamic changes in the distances between channel-gating residues during the conformational transitions of TMEM16F and XKR8, as well as the related structural changes in XKR8 after C-terminal cleavage; the heatmaps show the changes in residue contacts along the conformational transition pathways; and the structural diagrams indicate the spatial positions of the corresponding conformational transition pathways in the protein structures. (e) Scoring results of the supervision module for this independent research task, including the integrated evaluation of indicators such as strategy design, execution blueprint, report quality, and transfer capability.

Finally, Agent 4 (Supervision & Rating) evaluated the full-process performance of this independent research task according to a unified evaluation standard, yielding Plan quality = 92/100, Execution blueprint quality = 84/100, Report quality = 87/100, and Transfer quality = 94/100 (Figure 4e). These results indicate that the trained MDAgent is no longer only capable of generating executable plans for complex membrane-protein research tasks, but can also define key gating sites based on the literature, organize comparative simulation conditions, extract conformational transition pathways, and produce research reports with mechanistic interpretive power, while also demonstrating high stability in cross-task transfer capability.

## 3. Discussion

The core bottleneck of molecular dynamics research has never been merely whether simulation commands can be executed automatically, but whether a real biological problem can be effectively translated into an executable, interpretable, and reviewable computational research workflow[18–25]. In actual research, what researchers face is not a single software operation problem, but a continuous process of scientific decision-making: they must integrate the literature background, methodological applicability boundaries, and specific research objectives to complete system design, sampling-strategy selection, analysis-path organization, as well as result interpretation and quality assessment. Based on this understanding, the MDAgent proposed in this study does not confine MD automation to workflow packaging at the software level, but instead unifies problem understanding, literature-driven strategy design, simulation execution, result integration, mechanistic interpretation, and quality control within the same multi-agent framework. Unlike traditional automation tools, which mainly address "how to complete a simulation more efficiently," MDAgent attempts to answer "how to formulate an executable, interpretable, and reviewable molecular simulation research plan around a specific biological problem." Therefore, the core contribution of this work lies not only in constructing a system capable of executing MD tasks, but more importantly in advancing AI-driven molecular simulation from single-step operational automation to a research organization process that more closely resembles a real scientific workflow.

From a methodological perspective, the primary gain of MDAgent comes from the multi-agent structure itself. Molecular dynamics research is not a single type of reasoning task, but rather involves multiple heterogeneous cognitive loads simultaneously, including problem modeling, method selection, system setup, simulation quality control, trajectory analysis, and mechanistic interpretation[26–28]. If these tasks are compressed into a single round of language generation, the result often tends to be incomplete plans, broken chains of evidence, or a disconnect between execution logic and result narration. In contrast, the multi-agent framework assigns literature retrieval and strategy design, simulation execution, report generation, and independent review to different roles, allowing different types of knowledge and judgment to be invoked at more appropriate stages, while achieving stage-to-stage linkage and traceable revision through structured intermediate outputs. Our results show that even without considering case-based learning, task decomposition and stage-wise organization alone can already significantly improve the completeness of research blueprints and report outputs. This indicates that what determines the capability of an AI simulation system is not only the language ability of the underlying model, but also the system design of "how the task is organized."

On this basis, case-based learning further constitutes the key feature that distinguishes this work from general prompt engineering or static knowledge enhancement methods. The results show that the improvement brought by case-based learning is not merely that it provides the model with more context, but that it enables the system to gradually accumulate transferable research experience across continuous tasks. After learning from multiple representative cases, what the system acquires is no longer a set of isolated local answers, but a reusable capability structure that can be transferred across tasks: it gradually learns how to truly perform a standard simulation task correctly, how to choose appropriate systems and sampling strategies around a specific problem, how to identify genuinely informative analytical metrics, and how to organize trajectory data into mechanistic narratives oriented toward biological questions. In other words, what case-based learning distills is not "what tasks have been done," but rather

"what should be done under what circumstances, which errors should be avoided, and to what extent certain conclusions can be supported." The growth of system capability does not come from a single longer prompt, but from the continuous compression, retrieval, and transfer of real task experience. This process bears a clear resemblance to the way domain researchers grow in real scientific research. Researchers' capabilities usually do not come from the mechanical invocation of static rules, but from the gradual accumulation of workflow norms, method-selection experience, failure-repair strategies, problem-oriented analytical templates, and result-presentation styles in the course of continuously solving specific problems. What is reflected in the continuous case-based learning of MDAgent is precisely a similar path of capability evolution, which is also consistent with recent observations from industrial practice on agent engineering.

Another important finding of this study is that case-based learning drives the system to evolve from a "metric calculator" into a "mechanistic interpreter." In many existing automated workflows, simulation outputs often remain at the numerical level of RMSD, RMSF, hydrogen-bond counts, PMF, or several distance parameters, while whether these results can truly answer the research question still requires additional interpretation and reorganization by the researcher. The improvement of MDAgent lies in the fact that it gradually learns to organize analytical chains around scientific questions: in protein–ligand systems, it no longer stops at describing stability, but further connects the results to catalytic-site locking and restriction of flexible regions; in problems related to lipid environments, it no longer merely reports density or RDF differences, but instead attempts to formulate state-dependent lipid recruitment models; in transporter and efflux pump cases, it also no longer relies only on a single energy barrier or average distance, but further analyzes gating networks and interchain cooperative relationships. This indicates that what the system has acquired is not the ability to "calculate more parameters," but rather the ability to "determine which parameters are truly relevant to the current scientific question and, on that basis, organize mechanistic interpretation." This point is particularly critical in real-world usage scenarios for experimental scientists, because what they truly need is often not more computational output, but a reliable explanation of "what these results mean."

Overall, this study shows that the significance of MDAgent does not lie in replacing any single software module, but in reorganizing the entire process of molecular dynamics research "from question to conclusion." The multi-agent architecture provides task division and workflow control, case-based learning provides experience accumulation and cross-task transfer, and the structured review mechanism provides internal constraints on result quality. Together, these three components make the system no longer merely an automation tool capable of executing simulation commands, but one that begins to demonstrate the ability to organize a complete MD research process around biological questions. Together, these three components make the system no longer just an automation tool for executing simulation commands, but one that begins to possess the capability to organize a complete research workflow around scientific questions. In this sense, this work not only addresses the core challenge in molecular dynamics automation, but also proposes an AI collaboration paradigm that may be applicable to a broader range of computational research fields.

## 4. Materials and Methods
### 4.1. Overview of the MDAgent framework

We constructed MDAgent, a multi-agent system for end-to-end molecular dynamics research tasks. Unlike traditional MD automation workflows, which mainly focus on input-file generation, simulation-command invocation, or post-processing scripts, MDAgent organizes a complete computational study into continuous and connected scientific workflow units, including scientific problem understanding, literature retrieval and strategy design, simulation execution and analysis, result integration and biological narration, as well as final independent review and quality control, thereby achieving a complete closed loop from "posing a question" to "forming interpretable research conclusions." The underlying execution of the system is built on OpenClaw; at the start of each task, the framework dynamically instantiates a group of temporary agents around the current research problem and assigns each agent an independent workspace, allowing different stages to maintain role isolation, clear inputs, and traceable outputs while sharing the same overall objective. All intermediate results are stored in the form of structured files, thereby forming a task execution chain that is traceable, reviewable, and reproducible.

In a standard run, MDAgent consists of four functionally differentiated roles. Agent 1 is the research agent, which is responsible for translating the user's question into a clearly defined scientific problem and, in combination with domain background, extracting relevant literature clues, candidate methods, key control variables, control designs, and analysis metrics, ultimately outputting a structured research plan that can be passed downstream for execution. Agent 2 is the MD agent, which organizes system construction, parameter selection, simulation execution, and preliminary analysis based on the upstream plan; for different task types such as standard all-atom MD, membrane protein systems, coarse-grained simulation, umbrella sampling, metadynamics, or REMD, the system explicitly writes method selection, key parameters, expected outputs, quality control checkpoints, and rework conditions into the execution blueprint to ensure that subsequent runs are reviewable. Agent 3 is the report agent, which is responsible for integrating simulation outputs with the literature background to generate an analysis report for scientific use; this report includes not only numerical metrics and figure suggestions, but also biological interpretations of stability changes, interfacial interactions, gating mechanisms, conformational transitions, or free energy features. Agent 4 is the review agent, which independently examines whether the outputs of the preceding stages contain methodological errors, broken chains of evidence, insufficient mechanistic interpretation, excessive extrapolation of results, or incomplete figure design, and provides the final review conclusion in the form of a structured score.

The four roles are connected through explicit stage-wise input–output interfaces rather than loose conversational chaining. Specifically, the system generates standardized JSON artifacts at each stage, such as literature_plan in the literature and planning stage, md_execution in the simulation stage, report in the reporting stage, and report_review in the final review stage. On the one hand, these structured files serve as the direct input for downstream agents; on the other hand, they also constitute part of the complete experimental record, allowing the task to be paused, reviewed, or rerun at any stage. The review agent not only provides an overall score at the end of the workflow, but also performs stage-gating review after Agent 1 and Agent 2, respectively: if it detects that the research plan is insufficient, key parameters are missing, case transfer is inappropriate, or the execution blueprint is incomplete, it will return required_repairs and next_stage_focus to the upstream stage, requiring revision before proceeding to the next step.

On top of the multi-agent framework, we introduced case-based learning. After each MD task is completed, the system extracts the task type, retrieval path, planning logic, key parameters, analytical focus, and failure-repair information to form standardized case cards; at the same time, it distills operational workflows, quality-control checkpoints, and error-repair strategies into skill cards. When a new task arrives, the system first identifies the system type, research objective, and analysis requirements, and then retrieves a small number of highly relevant entries from the case library and skill library, which are injected into the independent workspaces of the temporary agents according to their roles. All agents output results in a unified structure to ensure the stability of information transfer, automatic review, and cross-task transfer.

## 4.2. Case-based learning for continual improvement

To evaluate and train the multi-agent system to progressively acquire molecular dynamics research capabilities approaching those of domain researchers over continuous tasks, we constructed nine representative case-based learning tasks. These cases collectively cover soluble proteins, nucleic acid complexes, membrane proteins, GPCRs, transmembrane small molecules, and multimeric ultra-large complexes at the system level; at the methodological level, they cover standard all-atom MD, coarse-grained MD, umbrella sampling, steered MD, replica-exchange MD, and well-tempered metadynamics[5,32–39]; and at the analytical level, they cover multiple paradigms including stability, interfacial contacts, interaction energy, lipid density, conformational transition parameters, and free energy reconstruction. Therefore, they constitute a highly diverse learning set for evaluating cross-task transfer and capability growth.

In terms of knowledge organization, we further extracted two types of reusable resources from each case. The first type is case memory, which records the background of the scientific problem, key points of system modeling, critical decision nodes, common failure modes, boundaries of result interpretation, and their applicable scenarios. The second type is operational skill, which encodes procedural methodological knowledge, including parameter settings, quality control checklists, trajectory post-processing workflows, key analysis commands, and error-repair strategies. Rather than simply storing historical answers, this case knowledge is compressed into shared units that are retrievable, composable, and transferable, so that they can be used for similar-case matching and runtime invocation in subsequent tasks, thereby supporting capability accumulation and methodological transfer across continuous tasks.

The simulation systems of the nine cases were as follows:

(1) All-atom molecular dynamics case of lysozyme. PDB 1AKI was used as the initial protein template, and an aqueous system was constructed under CHARMM36/TIP3P and 0.15 M NaCl conditions. After energy minimization, 50 ps NVT, and 50 ps NPT, a 200 ns production simulation was performed[2].

(2) HIV-1 protease–inhibitor complex case. The HIV-1 protease homodimer–DMP450 complex in PDB 1DMP was used as the reference structure, and an aqueous system was constructed under CHARMM36/TIP3P and 0.15 M NaCl conditions. After energy minimization, 50 ps NVT, and 50 ps NPT, a 200 ns production simulation was performed[6].

(3) p53–DNA complex case. PDB 3Q05 was used as the structural template for the DNA-bound p53 complex. Under CHARMM36m/TIP3P and 0.15 M NaCl conditions, a system containing three p53 DNA-binding domain chains and a 21 bp DNA response element was simulated for 200 ns.

(4) IAA blood–brain barrier permeation case. Deprotonated indole-3-acetate (IAA) was described using CHARMM36m + CGenFF, and 16 umbrella sampling windows were set up along the membrane normal in a CG brain-membrane lipid model composed of DOPC/POPE/PSM/CHL1, with each window simulated for 50 ns under a restraint constant of 1000 kJ mol$^{-1}$ nm$^{-2}$, and the transmembrane PMF was reconstructed using WHAM[40].

(5) GLUT1 glucose transport case. In this case, the inward-open conformation used the experimental structure PDB 4PYP as the starting template, whereas the outward-facing conformation was obtained by homology modeling with MODELLER using the outward-open crystal structure of Escherichia coli XylE, PDB 4GBZ, as the template. Following the workflow of Liu et al. (2024), a total of 100 outward-opening models were constructed, and conformations for subsequent simulations were selected based on structural rationality assessment. Both conformations were subjected to 23 umbrella sampling windows, with a total sampling time of 690 ns for each[40].

(6) NTSR1 lipid fingerprint case. PDB 5T04 was used as the structural reference for NTSR1. The receptor was converted into a Martini 2.2 coarse-grained model, and a 20 μs trajectory in a POPC/POPE/CHOL/SM + POP2 (PIP2) membrane environment was analyzed[32].

(7) **D2R REMD case.** Temperature replica-exchange simulations were performed to compare the inactive and active states of the dopamine D2 receptor, using PDB 6CM4 (DRD2–risperidone complex) as the structural reference for the inactive state and PDB 6VMS (agonist-bound activated DRD2–Gi complex) as the reference for the active state. The temperature range was 415–450 K, with 18 replicas, and the total sampling time was approximately 180 ns for each[3].

(8) MT1–melatonin metadynamics case[41,42]. In a CHARMM36 membrane system, a complex environment of the MT1 receptor (PDB: 6ME4) and melatonin was constructed, with membrane composition of POPC/POPE/OSM/NSM + CHL1. Well-tempered metadynamics was used, with the biased variable defined as the distance between the center of mass of the ligand and the center of mass of the receptor core region. The parameters were Gaussian height 1.5 kcal/mol, sigma 0.35 Å, bias factor 15, and a total length of 200 ns[38].

(9) AdeB efflux pump case. Based on the trimeric experimental structure of Acinetobacter baumannii AdeB, PDB 7KGI, simulations were performed for 500 ns in a CHARMM36 and POPE/POPG membrane environment[43].

### 4.3. Molecular dynamics case of scramblase conformational transition

Using PDB 8B8Q of TMEM16F and PDB 8XEJ of XKR8 as the initial structural templates, membrane-protein molecular dynamics systems were constructed in explicit membrane/water/ion environments. The membrane composition was the standard mammalian cell membrane POPC/POPE/OSM/NSM + CHL1. All systems were parameterized under the CHARMM36 force field and the TIP3P water model, and 0.15 M NaCl was added to maintain electroneutrality and near-physiological salt concentration. For TMEM16F, its $Ca^{2+}$-bound close conformation was retained in order to observe the evolution from the closed state toward an open-like conformation in the context of calcium binding; for XKR8, the C-terminal amino acids after D355 were removed during the modeling stage to simulate the activated state in which the autoinhibitory tail is released after caspase-3 cleavage. After energy minimization and short NVT and NPT equilibration, the systems entered production simulations.

**4.4. Baseline comparison and ablation experiment design**

To systematically evaluate the impact of multi-agent organization and the case-based learning (CBL) enhancement mechanism on the quality of molecular dynamics research tasks, we constructed a unified framework for baseline comparison and ablation evaluation. Specifically, each method was required to generate, around the same set of biomolecular simulation tasks, a set of research outputs with literature support, execution feasibility, and a closed loop of biological interpretation, rather than merely providing fragmented parameter suggestions or generic descriptions.

Within this evaluation framework, each benchmark task required the method to generate three types of standardized outputs:

(i) a literature-grounded simulation plan, used to describe the scientific question, method selection, key simulation strategies, and the basis for major parameter choices;

(ii) an MD execution blueprint, used to describe system construction, equilibration workflow, production-stage design, quality control checkpoints, and rework conditions;

(iii) a biological report draft, used to organize the expected analysis metrics, chains of evidence, figure suggestions, and mechanistic interpretation.

Through this design, we were able to uniformly compare whether different methods can formulate a molecular simulation research workflow that is scientifically sound, executable, and capable of mechanistic narration, without relying on the convergence results of real long-timescale trajectories. In this study, five method configurations were established to distinguish the independent contributions of task organization and knowledge enhancement sources. The first was Single-Agent LLM, in which a single large language model directly completed the entire process from problem understanding and plan generation to report organization, without adopting multi-agent division of labor or accessing any external case memory or skill knowledge. The second was Multi-Agent w/o CBL, which adopted the same multi-stage multi-agent workflow as the full system but did not introduce case memory or skill knowledge, and was used to evaluate whether stage-wise organization and task decomposition alone could improve output quality. The third was Memory Only, which accessed only case memory on top of the multi-agent workflow, without introducing skill knowledge, and was used to evaluate the value of historical case transfer itself. The fourth was Skill Only, which accessed only skill knowledge within the multi-agent workflow, without introducing case memory, and was used to evaluate the independent impact of methodological skills on the quality of research blueprints. The fifth was Full MDAgent + CBL, which simultaneously accessed case memory and skill knowledge and integrated both into different stages through a unified expert-routing mechanism; this was the complete method of the present study. Except for Single-Agent LLM, the other four methods all shared the same four-stage multi-agent organizational framework: Agent 1 was responsible for scientific problem understanding, literature synthesis, and simulation plan design; Agent 2 was responsible for generating the MD execution blueprint, including system construction, simulation-stage setup, and analysis workflow planning; Agent

3 was responsible for biological interpretation and report organization; and Agent 4 was responsible for final review, cross-stage consistency checking, and structured scoring.

To evaluate the generalization ability of the methods under cross-task conditions, we constructed a benchmark task set covering multiple biomolecular simulation scenarios, comprising a total of 10 benchmark cases: KRas, CFTR, P-gp, SARS-CoV-2 Mpro, NMDA receptor, AQP4, PARP1, ANT1, TLR4–MD2, and CCR5.

Specifically, the KRas task corresponds to the KRas G12D–switch-II pocket inhibitor system, with the core focus on mutation-induced conformational plasticity, pocket stability, and identification of the local binding microenvironment. This task tests whether a method can organize a reasonable protein–ligand equilibrium and local flexibility analysis in the context of an oncogenic mutation, rather than simplifying it into an ordinary binding problem.

The CFTR task focuses on the relationship between correctors and membrane-protein state stability in the context of the F508del mutation. Its difficulty lies in simultaneously handling the mutation, membrane environment, correctors, and conformational compensation mechanism, and mainly tests whether a method can organize a reasonable workflow in a large membrane-protein system and explain local structural recovery and interfacial stability.

The P-gp task concerns the efflux process of doxorubicin by P-glycoprotein, representing a typical problem of membrane transporter modeling and reaction-coordinate design. This task requires the method not only to build the complex system, but also to define a reasonable sampling coordinate for the efflux pathway or transmembrane migration, and therefore particularly tests the executability of the umbrella sampling scheme.

The SARS-CoV-2 Mpro task is oriented toward enhanced-sampling analysis of the main protease and a covalent inhibitor. This system simultaneously involves binding conformation, active-site geometry, and the pre-reactive organizational state, and is therefore suitable for testing whether a method can recognize the limitations of ordinary MD and introduce enhanced-sampling strategies such as metadynamics when needed.

The NMDA receptor task corresponds to REMD-based analysis of activation-related behavior in the ligand-binding domain, and is essentially a problem of conformational heterogeneity and activation mechanism within a multistate conformational space. This task requires the method to recognize conformational heterogeneity and to convert REMD sampling results into interpretations of mechanisms related to the open/closed states.

The AQP4 task investigates the coarse-grained lipid fingerprint of the relationship between the AQP4 water channel and the lipid environment. Its focus is not on a single ligand, but on the lipid composition around the protein, enrichment preferences, and local membrane-environment remodeling; therefore, it mainly tests whether the method has the ability to choose coarse-grained resolution according to the scientific question.

The PARP1 task focuses on the stability and interfacial mechanism of the complex involved in DNA damage recognition by the PARP1 zinc finger. Unlike ordinary protein–small molecule systems, this task places greater emphasis on the protein–DNA interface, damage-site recognition, and local structural adaptation, and therefore imposes higher requirements on mechanistic reporting and interface interpretation.

The ANT1 task centers on the nucleotide transport process of the mitochondrial ADP/ATP translocase and represents a typical transporter free energy pathway problem. This task examines whether the method truly understands the transport problem, and whether it can construct a reasonable reaction coordinate around the transmembrane pathway and organize free energy analysis, rather than remaining at the level of conventional MD observation.

The TLR4–MD2 task concerns the stability of the dimer interface induced by lipid A and represents an interface-organization problem in immune receptor complexes. Its difficulty lies in simultaneously handling the protein–protein interface, auxiliary ligands, and the context of receptor activation, and is therefore suitable for evaluating plan completeness and the ability to generate mechanistic interpretations at the complex level.

The CCR5 task is oriented toward coarse-grained analysis of cholesterol hotspots associated with CCR5 and represents a problem of cooperative regulation between membrane receptors and the lipid environment. This task requires the method to connect membrane-lipid hotspots, receptor surface regions, and coarse-grained sampling, and to form a clear framework for hotspot definition and functional interpretation.

All outputs were scored using the same structured review workflow. The scoring system contains four primary dimensions: plan quality, execution blueprint quality, report quality, and Adaptation & Transfer.

(1) Plan quality: This dimension is used to evaluate whether the method selection is reasonable, whether the key parameters have a scientific basis, and whether the overall research plan is complete. It contains three secondary fields: method_selection_accuracy, which evaluates whether the method selection is correct and aligned with the research task; parameter_rationality, which evaluates whether the key parameters are reasonable and scientifically justified; and plan_completeness, which evaluates whether the research plan forms a closed loop and has execution completeness. The average of these three items is taken as plan_mean.

(2) Execution blueprint quality: This dimension is used to evaluate whether the execution blueprint is practically deliverable, including whether system preparation is clearly defined, whether the workflow connection forms a closed loop, and whether quality control and rework checkpoints are clearly specified. It contains four fields: task_completion_rate, which measures the completeness of the execution blueprint; first_pass_success_rate, which measures the success of first-pass drafting or first-pass delivery; rework_count, which is the number of rework rounds and is a non-negative integer; and convergence_quality, which evaluates the quality of convergence and quality-control design. The average of these four items is taken as execution_mean.

(3) Report quality: This dimension is used to evaluate whether the report has scientific correctness, depth of mechanistic interpretation, content completeness, and figure organization capability. It contains five secondary fields: scientific_correctness; mechanistic_depth; content_richness; figure_completeness; and evidence_completeness. The average of the five items is taken as report_mean.

(4) Transfer quality: This dimension is used to evaluate whether the system can accurately transfer relevant historical experience to the current task and perform targeted adaptation rather than mechanical reuse. It contains four secondary fields: case_selection_precision; adaptation_depth; cross_stage_traceability; and novel_task_value_add. The average of these four items is taken as transfer_mean.

**4.5. Molecular dynamics simulations**

All conventional molecular dynamics simulations were performed using GROMACS 2021.2, and the initial models for all cases were built based on experimental structures[1]. Protein and nucleic acid coordinates were obtained from the corresponding PDB entries. Before system construction, crystallographic heteroatoms, low-occupancy alternative conformations, and components without parameters were removed, retaining only the proteins, nucleic acids, key structural waters, and necessary cofactors required for the study. Aqueous systems were constructed using the CHARMM36m/TIP3P protocol within a local GROMACS workflow, and salt ions were added to achieve an ionic strength of approximately 0.15 M while neutralizing the system charge and mimicking physiological conditions. For protein–ligand systems, small-molecule ligands were taken from the co-crystal structures, and their parameters were generated using a small-molecule parameterization workflow compatible with CHARMM36. For membrane-protein systems, the transmembrane orientation of the protein was determined according to OPM/PPM or the original literature, and the membrane environment was built using CHARMM36 native lipid templates[44,45]. For systems using the coarse-grained route, the Martini 3 force field was used consistently for local modeling. Using the atomistic model as input, coarse-grained coordinates and topologies were generated with martinize2 according to the Martini 3 mapping rules[32,33]. All systems were ultimately organized into topology, coordinate, index, and mdp files directly usable in GROMACS, and successful generation of a .tpr file by gmx grompp was taken as the criterion for completion of system construction. GROMACS input files were generally prepared in four stages: energy minimization, NVT, NPT, and production simulation, and were further extended in enhanced-sampling tasks to include pulling, umbrella sampling, replica exchange, or metadynamics inputs. Parameter settings followed standard practice for CHARMM36/CHARMM36m in GROMACS. All systems employed three-dimensional periodic boundary conditions. Energy minimization used the steepest descent algorithm, with the maximum force convergence threshold set to 1000 kJ mol$^{-1}$ nm$^{-1}$. For membrane-protein systems, after minimization, multistage equilibration was carried out according to the standard CHARMM-GUI workflow. In the early equilibration stages, positional restraints or dihedral restraints were applied to proteins, ligands, lipid headgroups, and relevant dihedral angles, and the restraint strength was gradually reduced in subsequent steps until entering the unrestrained production simulation. For soluble proteins, protein–ligand systems, and protein–DNA systems, short NVT and NPT pre-equilibration was usually performed before entering the production simulation. Production simulations were all carried out under the NPT ensemble, with the temperature maintained at 313.15 K by the Nosé–Hoover thermostat and the pressure controlled at 1 bar by the Parrinello–Rahman barostat; membrane systems used semi-isotropic pressure coupling, whereas non-membrane systems used isotropic pressure coupling[46–48]. Long-range electrostatic interactions were calculated using the particle mesh Ewald (PME) method. Short-range Coulomb and van der Waals interactions generally used a cutoff radius of 1.2 nm, and van der Waals interactions were treated with a smooth cutoff scheme compatible with CHARMM36m[49,50]. All bonds involving hydrogen atoms were constrained using LINCS, and water geometry was constrained using SETTLE, thereby allowing an integration time step of 2 fs[50,51]. Before formal analysis, all production trajectories underwent standard post-processing, including removal of across-box artifacts introduced by periodic boundary conditions, centering, and necessary least-squares fitting, to ensure the reliability of RMSD, interfacial contact, gating distance, lipid density, and free energy analyses.

For problems that require access to conformational space, transmembrane pathways, or free energy landscapes that are difficult to adequately cover by conventional MD, this study further introduced enhanced sampling methods, including umbrella sampling, steered molecular dynamics (SMD), temperature replica-exchange molecular dynamics (T-REMD), and well-tempered metadynamics[3,36–38,52]. In umbrella sampling tasks, initial conformations distributed along a selected reaction coordinate were first generated through pre-pulling or pathway construction, after which multiple sampling windows were defined, and harmonic positional restraints were applied in each window for independent sampling; finally, WHAM was used to integrate the sampling results from all windows and reconstruct the PMF curve. SMD was used to obtain nonequilibrium pathway information, identify potential resistance sites and candidate transport channels, and when a thermodynamic free energy profile needed to be constructed, SMD was usually combined with umbrella sampling, with the former used for pathway exploration and the latter for near-equilibrium free energy reconstruction. For REMD tasks, this study adopted a temperature replica-exchange scheme, in which multiple replicas were run in parallel under a preset temperature gradient, and exchanges between adjacent temperatures were attempted at a fixed frequency according to the Metropolis criterion, thereby improving the ability of the system to cross local energy barriers and sample multiple conformational states. In result analysis, the focus was placed on comparing the distributions of key conformational parameters under different states, the sampling coverage after replica exchange, and the statistical separation of structural markers, rather than relying on a single representative conformation to draw conclusions. For well-tempered metadynamics, the bias potential was applied in GROMACS through the PLUMED plugin, enhancing the exploration of low-probability state space by periodically depositing Gaussian potentials along predefined collective variables (CVs)[36,37]. In such tasks, in addition to reconstructing the free energy surface, we further examined the boundaries of interpretation by combining CV time distributions, state occupancy, bias-potential evolution, and trajectory revisitation behavior, so as to distinguish between "pathway and state information that is already reliably supported by the current sampling" and "quantitative free energy differences that still require more sufficient sampling for validation."

**4.6. Simulation analysis**

Most structural properties, dynamic properties, interaction parameters, and free-energy-related analyses were performed using built-in GROMACS tools, with the corresponding tool names indicated in parentheses after each parameter where applicable. For tasks requiring customized trajectory filtering, chain-specific statistics, coarse-grained lipid distribution calculation, REMD distribution comparison, or metadynamics free energy surface reconstruction, additional tools including MDAnalysis, pytraj, NumPy, Matplotlib, and PLUMED were also used[36,37,53].

For the scramblases: for the TMEM16F (8B8Q) and XKR8 (8XEJ) trajectories, we first performed standardized preprocessing through periodic boundary condition correction, molecular stitching, system centering, and conformational alignment, and combined thermodynamic quantities, RMSD, RMSF, and Rg to confirm that the systems had reached an analyzable stable state. At the mechanistic analysis level, gating-residue distances were used as the primary parameters, together with residue contact differences and conformational transition pathway analysis, to characterize function-related conformational transitions. For TMEM16F, I515–K616 was used as the transport-pocket gating metric, whereas for XKR8, W45–I152 in chain B was used as the transport-pocket gating metric. The gating distance was then used as the direct readout of function-related conformational change, and the trajectories were divided into two representative states, namely the initial conformation and the terminal conformation. On this basis, residue contact differences between the two states were compared on the one hand, while dynamic changes in residue contact frequency before and after conformational transition were quantified on the other. Only contact changes that simultaneously satisfied the criteria of being "significantly associated with gating transition," "forming a spatially continuous propagation relationship," and "being able to connect the activation starting point with the functional output end" were included within the scope of conformational transition pathway identification.

For the case-based learning examples: (1) All-atom molecular dynamics case of lysozyme: For the 1AKI lysozyme trajectory, periodic boundary condition correction, molecular stitching, and centering were first performed to eliminate the interference of across-box artifacts on structural statistics. The backbone root mean square deviation, RMSD (gmx rms), was then calculated to evaluate overall structural stability; the per-residue root mean square fluctuation, RMSF (gmx rmsf), was calculated to identify highly flexible regions; and the radius of gyration, Rg (gmx gyrate), was calculated to characterize conformational compactness. At the same time, basic thermodynamic quantities, including temperature, pressure, density, total energy, and potential energy (gmx energy), were extracted to examine the equilibration state of the system. On this basis, a covariance matrix was further constructed and principal component analysis was performed (gmx covar), with eigenvector projections used to describe the major conformational motions (gmx anaeig), and the free energy landscape was reconstructed based on the two-dimensional RMSD–Rg distribution in order to determine whether the system exhibited a single dominant stable basin or contained potential metastable states.

(2) HIV-1 protease–inhibitor complex case: For the HIV-1 protease–DMP450 inhibitor complex, the overall stability of the protein was first analyzed, including the backbone RMSD (gmx rms), local residue flexibility RMSF (gmx rmsf), and radius of gyration Rg (gmx gyrate). At the same time, ligand binding stability was examined as a key focus, including the ligand heavy-atom RMSD relative to the crystal binding conformation, the distance variation between the catalytic residue Asp25 and the key carbonyl atom of the ligand (gmx distance), and the number and occupancy of hydrogen bonds between the ligand and the active site (gmx hbond). In addition, the flexibility of flap-region residues 48–54 was specifically quantified to characterize whether the inhibitor constrained the flap in a closed conformation. For nonbonded interactions, after constructing custom indices for the protein and ligand, the electrostatic Coulomb and van der Waals Lennard–Jones components between key residues and the ligand were further extracted and calculated (gmx energy), thereby enabling energy-based interpretation of the "locking" mechanism by which the inhibitor occupies the hydrophobic pocket, locks the catalytic residues, and restricts flap motion.

(3) p53–DNA complex case: For the p53–DNA complex trajectory, PBC correction and trajectory continuity processing were first performed, followed by RMSD (gmx rms) analysis to evaluate the overall stability of the complex and conformational fluctuations at different time stages, and RMSF (gmx rmsf) analysis to characterize the flexibility distribution of different protein residues. For protein–DNA interactions, the number of hydrogen bonds between the protein and the DNA bases or backbone and their time evolution were calculated (gmx hbond), and interfacial contact residues, persistent contact patterns between hotspot charged residues and DNA, were further quantified to identify interface-dominant residues and potential mutation-sensitive sites. In result interpretation, the proportion of stable versus unstable states, the fluctuation range of interfacial hydrogen-bond numbers, and the retention of interfacial contacts were combined to evaluate whether the complex maintained stable recognition, whether dynamic "breathing" behavior occurred, and how interfacial stability was related to the functions of classical hotspot residues such as R248 and R273.

(4) Drug blood–brain barrier crossing case: For the IAA transmembrane permeability task, multiple sampling windows were constructed by umbrella sampling along the membrane-normal reaction coordinate, and the potential of mean force (PMF) curve was reconstructed using WHAM (gmx wham). The analysis included the position of the free energy minimum, the position of the highest energy barrier, the total transmembrane energy barrier, the sampling coverage and overlap between different windows, and the trend of the free energy curve across the membrane headgroup region and hydrophobic core region. In addition, combined with the free energy gradient or local resistance distribution, the major source of resistance encountered by the charged carboxylate group when entering the hydrophobic membrane region was analyzed, and on this basis it was assessed whether its passive diffusion capability was limited. In interpreting the results, attention was paid not only to the magnitude of the total energy barrier, but also to whether the free energy minimum was located in the membrane headgroup region and whether a local "trap" binding site was formed.

(5) GLUT1 transporter case: The analysis of the GLUT1 transport mechanism combined both equilibrium free energy analysis and nonequilibrium mechanical analysis. For the outward-open and inward-open conformations, umbrella sampling was used to sample substrate displacement along the transport pathway, and WHAM was used to reconstruct the PMF (gmx wham), thereby obtaining the total transport energy barrier, the positions of free energy peaks and valleys, and their distribution along the pathway. At the same time, local analyses were also performed on key gating residues and the gating network, including distance changes between key residue pairs, as well as the average hydrogen-bond number, occupancy, and time evolution of the Q161–W388–H160 hydrogen-bond network. Finally, through joint analysis of the PMF, pulling force, and key gating contacts, it was determined which side of the gating conformation constituted the major rate-limiting step.

(6) NTSR1–GPCR coarse-grained membrane system case: For the coarse-grained membrane simulation of NTSR1, the main focus was on analyzing the spatial distribution of different lipid components around the receptor and their relative enrichment/depletion features. Specifically, the analyses included the radial distribution functions (RDFs) of lipids around the receptor, two-dimensional density maps of different lipid types, the positional distribution of the protein center of mass relative to the membrane bilayer, the z-coordinate range along the membrane normal, the minimum distance between the protein and the membrane surface, and the statistical abundance of different lipids within local regions surrounding the receptor. By comparing the RDFs and density distributions of lipids such as PIP2, CHOL, POPC, and POPE, we determined whether cholesterol hotspots, PIP2 enrichment, or depletion existed around the receptor, and further interpreted their relationships with the charge distribution on the intracellular surface of the GPCR, membrane-surface binding states, and potential Gq-coupled conformations.

(7) D2R REMD conformational transition case: For the replica-exchange simulations of the inactive and active states of the dopamine D2 receptor (D2R), the main analysis focused on changes in the classical conformational markers during GPCR activation. Specifically, the distance between key residues on the intracellular side of TM3 and TM6 was first used as the primary reaction parameter, and its mean, median, distribution width, and time evolution were quantified in the inactive and active trajectories. This was followed by further analysis of the radial outward displacement of TM6 and its movement along the membrane normal, in order to quantify the outward swing of TM6 and the downward movement of its intracellular end during receptor activation. In addition to geometric parameters, statistical tests and effect size calculations were also performed on the distance distributions in the two states, so as to evaluate the significance and biological relevance of the conformational differences between the inactive and active states.

(8) MT1–melatonin metadynamics case: For the well-tempered metadynamics simulation of the MT1 receptor with melatonin, the evolution trajectory of the collective variable over time, as recorded in the COLVAR file, was first analyzed to describe the process by which the ligand gradually approached from the extracellular solution and entered the receptor binding pocket. At the same time, the time-segment characteristics of the pre-binding exploration stage, the approach stage, and the stable binding stage were quantified, together with the fraction of time during which the ligand remained in the tightly bound state. For the biasing history, the cumulative deposition of Gaussian bias in the HILLS file was analyzed. The free energy analysis was mainly based on reconstruction of the free energy surface (FES/PMF) using PLUMED sum_hills, from which the positions of free energy minima, binding intermediates, metastable residence regions, and relative free energy differences across different reaction-coordinate intervals were extracted.

(9) AdeB multicomponent efflux pump case: For the AdeB trimeric efflux pump system, chain-specific analysis was performed separately for the three subunits PROA, PROB, and PROC. First, the time series of the distances between the key inner-gate residue pair L137–F612 and the key outer-gate residue pair R931–D407 were calculated for each subunit, and their mean values, fluctuation ranges, and state distributions were compared. Subsequently, the gradient differences and asymmetry indices of the inner-gate and outer-gate distances among the subunits were quantified in order to determine whether the three subunits occupied different conformational states. In addition, the correlations between gate distances, interchain conformational asynchrony, and the occupancy ratios of different gate states were also analyzed. Finally, through chain-specific distance distributions, gradient arrangements, and correlation analysis, the conformational dynamic basis of three-subunit cooperativity, asynchronous progression, and ratchet-like unidirectional efflux was inferred.

## Associated Content

## Author Information

### Corresponding Author

*E-mail: xukai.jiang@sdu.edu.cn

### Author Contributions

The manuscript was written through contributions of all authors. All authors have given approval to the final version of the manuscript.

### Notes

The authors declare no competing financial interest.

### Acknowledgement

This work was supported by the National Key Research and Development Program of China (2023YFC3403502), the National Natural Science Foundation of China (32301041, 32571437), the Shandong Excellent Young Scientists (Overseas) Fund Program (2023HWYQ-044), and the SKLMT Frontiers and Challenges Project (SKLMTFCP-2023-01).

**References:**

1. Abraham, M. J. *et al.* GROMACS: High performance molecular simulations through multi-level parallelism from laptops to supercomputers. *SoftwareX* **1–2**, 19–25 (2015).

2. Park, S.-J., Kern, N., Brown, T., Lee, J. & Im, W. CHARMM-GUI *PDB Manipulator*: Various PDB Structural Modifications for Biomolecular Modeling and Simulation. *Journal of Molecular Biology* **435**, 167995 (2023).

3. Suh, D. *et al.* CHARMM-GUI Enhanced Sampler for various collective variables and enhanced sampling methods. *Protein Science* **31**, e4446 (2022).

4. Park, S., Choi, Y. K., Kim, S., Lee, J. & Im, W. CHARMM-GUI Membrane Builder for Lipid Nanoparticles with Ionizable Cationic Lipids and PEGylated Lipids. *J. Chem. Inf. Model.* **61**, 5192–5202 (2021).

5. Huang, J. *et al.* CHARMM36m: an improved force field for folded and intrinsically disordered proteins. *Nat Methods* **14**, 71–73 (2017).

6. Lee, J. *et al.* CHARMM-GUI Input Generator for NAMD, GROMACS, AMBER, OpenMM, and CHARMM/OpenMM Simulations Using the CHARMM36 Additive Force Field. *J. Chem. Theory Comput.* **12**, 405–413 (2016).

7. Van Der Spoel, D. *et al.* GROMACS: Fast, flexible, and free. *Journal of Computational Chemistry* **26**, 1701–1718 (2005).

8. Thirunavukarasu, A. J. *et al.* Large language models in medicine. *Nat Med* **29**, 1930–1940 (2023).

9. Sallam, M. ChatGPT Utility in Healthcare Education, Research, and Practice: Systematic Review on the Promising Perspectives and Valid Concerns. *Healthcare (Basel)* **11**, 887 (2023).

10. Wang, L. *et al.* A Survey on Large Language Model based Autonomous Agents. *Front. Comput. Sci.* **18**, 186345 (2024).

11. Raiaan, M. A. K. *et al.* A Review on Large Language Models: Architectures, Applications, Taxonomies, Open Issues and Challenges. *IEEE Access* **12**, 26839–26874 (2024).

12. Obi, P. & Natesan, S. Membrane Lipids Are an Integral Part of Transmembrane Allosteric Sites in GPCRs: A Case Study of Cannabinoid CB1 Receptor Bound to a Negative Allosteric Modulator, ORG27569, and Analogs. *J Med Chem* **65**, 12240–12255 (2022).

13. Isu, U. H., Polasa, A. & Moradi, M. Differential behavior of conformational dynamics in active and inactive states of cannabinoid receptor 1 revealed by microsecond molecular dynamics simulation. *Biophysical Journal* **123**, 15a–16a (2024).

14. Guo, X., Li, F. & Zhang, F. Structural and dynamic mechanisms of cannabinoid receptors. *Biochemical Pharmacology* **244**, 117568 (2026).

15. Zhang, X. *et al.* Allosteric modulation and biased signalling at free fatty acid receptor 2. *Nature* **643**, 1428–1438 (2025).

16. Lu, S. *et al.* Activation pathway of a G protein-coupled receptor uncovers conformational intermediates as targets for allosteric drug design. *Nat Commun* **12**, 4721 (2021).

17. Ma, Z. *et al.* Transferable Expertise for Autonomous Agents via Real-World Case-Based Learning. Preprint at https://doi.org/10.48550/arXiv.2604.12717 (2026).

18. Roessner, R. A., Floquet, N. & Louet, M. Unveiling G-Protein-Coupled Receptor Conformational Dynamics via Metadynamics Simulations and Markov State Models. *J. Chem. Inf. Model.* **65**, 4630–4642 (2025).

19. Gomes, A. A. S. *et al.* Lipids modulate the dynamics of GPCR:β-arrestin interaction. *Nat Commun* **16**, 4982 (2025).

20. Conflitti, P. *et al.* Functional dynamics of G protein-coupled receptors reveal new routes for drug discovery. *Nat Rev Drug Discov* **24**, 251–275 (2025).

21. Picard, L. P. *et al.* Balancing G protein selectivity and efficacy in the adenosine A(2A) receptor. *Nat Chem Biol* **21**, 71–79 (2024).

22. Klein, F. *et al.* The SIRAH force field: A suite for simulations of complex biological systems at the coarse-grained and multiscale levels. *Journal of Structural Biology* **215**, 107985 (2023).

23. Yang, X. *et al.* Molecular mechanism of allosteric modulation for the cannabinoid receptor CB1. *Nat Chem Biol* **18**, 831–840 (2022).

24. Sandhu, M. *et al.* Dynamic spatiotemporal determinants modulate GPCR:G protein coupling selectivity and promiscuity. *Nat Commun* **13**, 7428 (2022).

25. Corradi, V. *et al.* Emerging Diversity in Lipid-Protein Interactions. *Chem Rev* **119**, 5775–5848 (2019).

26. Yu, D. *et al.* Application of the molecular dynamics simulation GROMACS in food science. *Food Research International* **190**, 114653 (2024).

27. Tan, X. *et al.* Decoding Electrochemical Processes of Lithium-Ion Batteries by Classical Molecular Dynamics Simulations. *Advanced Energy Materials* **14**, 2400564 (2024).

28. Molecular dynamics simulation in concrete research: A systematic review of techniques, models and future directions. *Journal of Building Engineering* **76**, 107267 (2023).

29. Barbhuiya, S. & Das, B. B. Molecular dynamics simulation in concrete research: A systematic review of techniques, models and future directions. *Journal of Building Engineering* **76**, 107267 (2023).

30. Bai, G. *et al.* Research advances of molecular docking and molecular dynamic simulation in recognizing interaction between muscle proteins and exogenous additives. *Food Chemistry* **429**, 136836 (2023).

31. Integration of Molecular Docking Analysis and Molecular Dynamics Simulations for Studying Food Proteins and Bioactive Peptides | Journal of Agricultural and Food Chemistry. https://pubs.acs.org/doi/full/10.1021/acs.jafc.1c06110.

32. Qi, Y. *et al.* CHARMM-GUI Martini Maker for Coarse-Grained Simulations with the Martini Force Field. *J. Chem. Theory Comput.* **11**, 4486–4494 (2015).

33. Arnarez, C. *et al.* Dry Martini, a Coarse-Grained Force Field for Lipid Membrane Simulations with Implicit Solvent. *J. Chem. Theory Comput.* **11**, 260–275 (2015).

34. Wassenaar, T. A., Pluhackova, K., Böckmann, R. A., Marrink, S. J. & Tieleman, D. P. Going Backward: A Flexible Geometric Approach to Reverse Transformation from Coarse Grained to Atomistic Models. *J. Chem. Theory Comput.* **10**, 676–690 (2014).

35. Qi, Y. *et al.* CHARMM-GUI PACE CG Builder for Solution, Micelle, and Bilayer Coarse-Grained Simulations. *J. Chem. Inf. Model.* **54**, 1003–1009 (2014).

36. Bonomi, M. *et al.* PLUMED: A portable plugin for free-energy calculations with molecular dynamics. *Computer Physics Communications* **180**, 1961–1972 (2009).

37. Tribello, G. A., Bonomi, M., Branduardi, D., Camilloni, C. & Bussi, G. PLUMED 2: New feathers for an old bird. *Computer Physics Communications* **185**, 604–613 (2014).

38. Barducci, A., Bussi, G. & Parrinello, M. Well-Tempered Metadynamics: A Smoothly Converging and Tunable Free-Energy Method. *Phys. Rev. Lett.* **100**, 020603 (2008).

39. Ligand Binding: Molecular Mechanics Calculation of the Streptavidin-Biotin Rupture Force. *Science* https://www.science.org/doi/10.1126/science.271.5251.997.

40. Nonphysical sampling distributions in Monte Carlo free-energy estimation: Umbrella sampling - ScienceDirect. https://www.sciencedirect.com/science/article/abs/pii/0021999177901218.

41. Stauch, B. *et al.* Structural basis of ligand recognition at the human MT1 melatonin receptor. *Nature* **569**, 284–288 (2019).

42. Johansson, L. C. *et al.* XFEL structures of the human MT2 melatonin receptor reveal the basis of subtype selectivity. *Nature* **569**, 289–292 (2019).

43. Lee, J. *et al.* CHARMM-GUI Membrane Builder for Complex Biological Membrane Simulations with Glycolipids and Lipoglycans. *J. Chem. Theory Comput.* **15**, 775–786 (2019).

44. Huang, J. *et al.* CHARMM36m: an improved force field for folded and intrinsically disordered proteins. *Nat Methods* **14**, 71–73 (2017).

45. OPM database and PPM web server: resources for positioning of proteins in membranes | Nucleic Acids Research | Oxford Academic. https://academic.oup.com/nar/article/40/D1/D370/2903396?login=true.

46. Nosé, S. A molecular dynamics method for simulations in the canonical ensemble. *Molecular Physics* **52**, 255–268 (1984).

47. Hoover, W. G. Canonical dynamics: Equilibrium phase-space distributions. *Phys. Rev. A* **31**, 1695–1697 (1985).

48. Parrinello, M. & Rahman, A. Polymorphic transitions in single crystals: A new molecular dynamics method. *Journal of Applied Physics* **52**, 7182–7190 (1981).

49. Kumar, N. & Garg, P. Probing the Molecular Basis of Cofactor Affinity and Conformational Dynamics of Mycobacterium tuberculosis Elongation Factor Tu: An Integrated Approach Employing Steered Molecular Dynamics and Umbrella Sampling Simulations. *J. Phys. Chem. B* **126**, 1447–1461 (2022).

50. Darden, T., York, D. & Pedersen, L. Particle mesh Ewald: An N·log(N) method for Ewald sums in large systems. *The Journal of Chemical Physics* **98**, 10089–10092 (1993).

51. Hess, B., Bekker, H., Berendsen, H. J. C. & Fraaije, J. G. E. M. LINCS: A linear constraint solver for molecular simulations. *Journal of Computational Chemistry* **18**, 1463–1472 (1997).

52. de Jong, D. H. *et al.* Improved Parameters for the Martini Coarse-Grained Protein Force Field. *J. Chem. Theory Comput.* **9**, 687–697 (2013).

53. MDAnalysis: A toolkit for the analysis of molecular dynamics simulations - Michaud-Agrawal - 2011 - Journal of Computational Chemistry - Wiley Online Library. https://onlinelibrary.wiley.com/doi/10.1002/jcc.21787.